\documentclass[a4paper,conference]{IEEEtran}
\IEEEoverridecommandlockouts

\usepackage{cite}
\usepackage{amsmath,amssymb,amsfonts}
\usepackage{algorithmic}
\usepackage{graphicx}
\usepackage{textcomp}
\usepackage{xcolor}
\usepackage{hyperref}
\usepackage{balance}
\usepackage{microtype}
\usepackage{tikz}
\usepackage{pgfplots}
\usepackage{pgfplotstable}
\usepackage{enumitem}
\pgfplotsset{compat=1.18}
\usepgfplotslibrary{statistics}
\usepgfplotslibrary{colorbrewer}

\usepackage[capitalize,nameinlink,noabbrev]{cleveref}
\crefname{figure}{Fig.}{Figs.}

\def\BibTeX{{\rm B\kern-.05em{\sc i\kern-.025em b}\kern-.08em
    T\kern-.1667em\lower.7ex\hbox{E}\kern-.125emX}}
\begin{document}


\title{Exploring React Library Related Questions on Stack Overflow: Answered vs. Unanswered}


\author{
  \IEEEauthorblockN{1\textsuperscript{st} Vanesya Aura Ardity}
  \IEEEauthorblockA{
    \textit{Informatics Engineering}\\
    \textit{Universitas Muhammadiyah Surakarta}\\
    Surakarta, Indonesia\\
    L200210170@student.ums.ac.id
  }
  \and
  \IEEEauthorblockN{2\textsuperscript{nd} Yusuf Sulistyo Nugroho}
  \IEEEauthorblockA{
    \textit{Informatics Engineering}\\
    \textit{Universitas Muhammadiyah Surakarta}\\
    Surakarta, Indonesia\\
    yusuf.nugroho@ums.ac.id
  }
  \and
  \IEEEauthorblockN{3\textsuperscript{rd} Syful Islam}
  \IEEEauthorblockA{
    \textit{Computer Science and Engineering Dept.}\\
    \textit{Gopalganj Science and Technology University}\\
    Gopalganj, Bangladesh\\
    syfulcse@gstu.edu.bd
  }
}

\maketitle

\begin{abstract}
React has been a popular JavaScript framework in contemporary web application development. 
Due to its high performance and efficiency, many web developers have made use of this framework. 
Although the react library offers many advantages, it is not without its challenges. When utilizing the react library, developers often run into problems for which they often seek solutions through question-and-answer forums, such as Stack Overflow (SO). However, despite its high popularity, numerous React-related questions on SO remain unanswered. 
Thus, this study aims to analyze the factors associated the question answerability and difficulty levels of React-related questions on SO. To facilitate our study, Exploratory Data Analysis (EDA) was applied to 534,820 questions, where they are filtered based on 23 React-related tags. We implemented a quantitative approach through text mining and statistical analysis. A logistic regression model was used to identify attributes associated with question answerability, while a simple linear regression model was employed to examine the correlation between user reputations and performance difficulty scores.
The results reveal that several attributes, such as the number of views, code snippet inclusion, number of lines of code, and user reputations positively affect the likelihood of a question answerability.
In contrast, the number of comments, question lengths, and the presence of image in the React-related questions reduce the probability of a question receiving responses from other users.
Further investigation indicates a negative correlation between user reputations and performance difficulty score (PD Score), where reputation increase corresponds to -0.092 reduction in PD score, signaling experienced users tendency to propose more complex technical inquiries.
This study provides insights into the discussion characteristics of technical question-and-answer platforms, such as Stack Overflow, that users need to consider the answerability factors when posting questions related to React.
\end{abstract}


\section{Introduction}
React, a popular JavaScript framework, was developed to address various challenges in web application development, particularly in terms of performance and efficiency. Throughout its evolution, React has outperformed other frameworks like Vue and Angular, thus dominating the front-end ecosystem~\cite{javeed2019performance}. 
React offers an advantage in its virtual DOM implementation, which not only improves overall application performance but also determines when a component needs to be reloaded based on occurring changes~\cite{chen2019front}. 

Despite the increasing popularity of React, however, many web developers also face challenges during their web development~\cite{samudio2022barriers}. They frequently encounter problems when coding to build web-based applications using React. 
To find solutions, many developers turn to online discussion forums, such as Stack Overflow (SO).
Although SO has become the largest Q\&A platform for programming topics to pose questions and search for technical solutions~\cite{xu2020makes},
however, a significant number of React-related questions remain unanswered.

Previous studies have analyzed various aspects of online discussion forums. For instance, a study on React discussion trends indicates a decline in activity regarding answering questions, providing comments, and giving ratings between 2020 and 2021~\cite{kurniaji2023preliminary}. In a similar context, a study of 2,322 SO posts related to Network Simulator (NS) found that users utilize SO as an implementation guide for NS models~\cite{islam2021network}. In addition, prior work on Python-related discussions on SO ~\cite{nugroho2024empirical} revealed that user reputation plays an important role in determining question characteristics and response likelihood. 
While these works provide valuable insights, they often focus on specific technologies without addressing both question answerability and difficulty. 
Addressing this gap, our study extends these works by focusing on React-related questions and analyzing the factors related to answerability and the complexity of questions.

Our findings reveal hidden patterns behind the answerability of React-related questions. Through quantitative analysis, we identify key factors associated with answerability, particularly the number of views, the number of comments, and the presence of code snippets. Furthermore, we explore the relationship between user reputation and question difficulty, uncovering a unique trend where higher technical expertise correlates with the formulation of more complex and in-depth questions in the context of React development.

\section{Research Method}

\begin{figure}
    \centering
    \includegraphics[width=.45\textwidth]{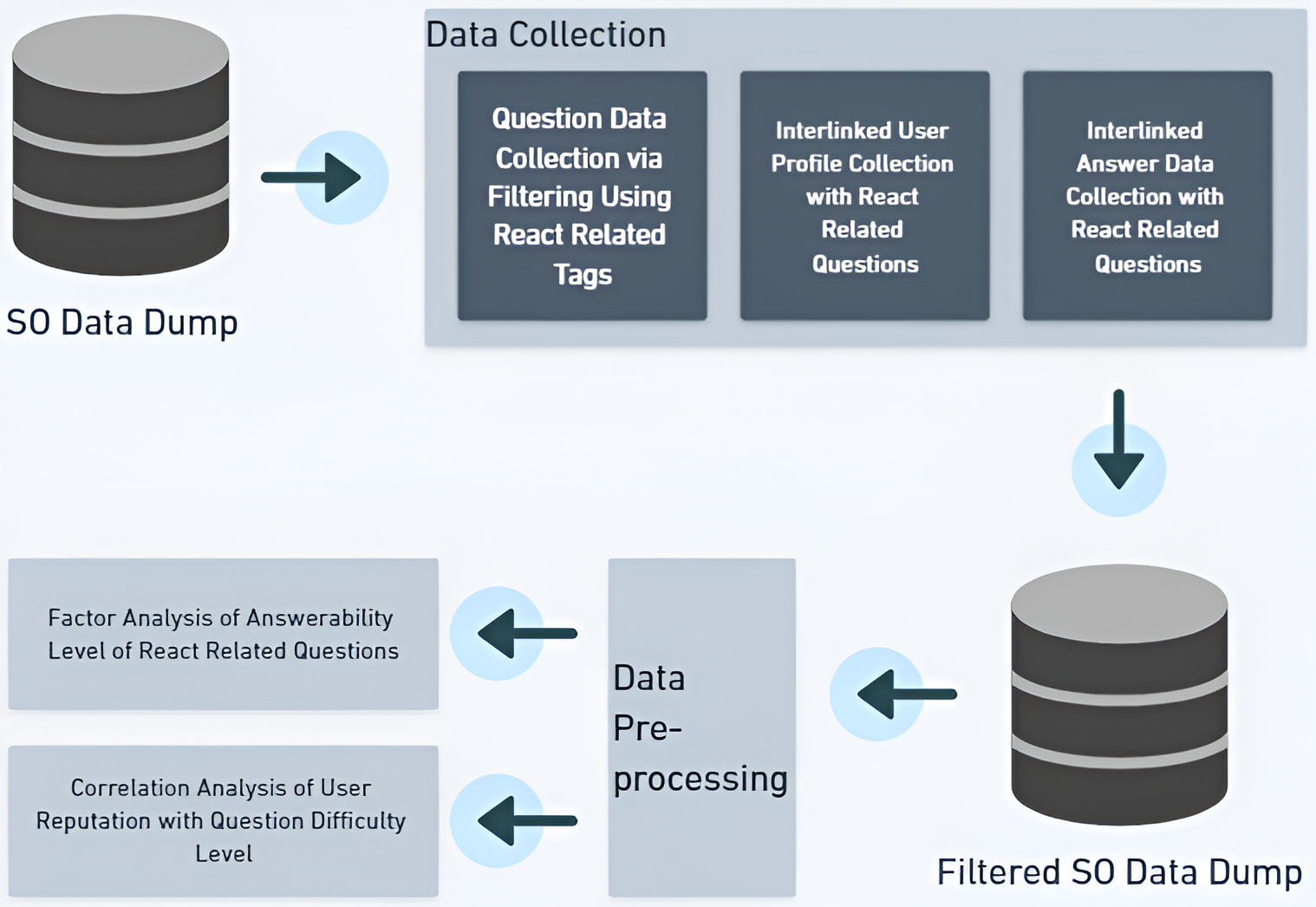}
    \caption{Overview of our research procedure.}
    \label{fig:overviewresearchsteps}
\end{figure}

In analyzing the complexity and the difficulty of React-related questions shared on SO, we applied the procedures from data collection to analysis, as illustrated in~\cref{fig:overviewresearchsteps}.

\subsection{Data Collection}

\begin{figure}
    \centering
    \includegraphics[width=.9\columnwidth]{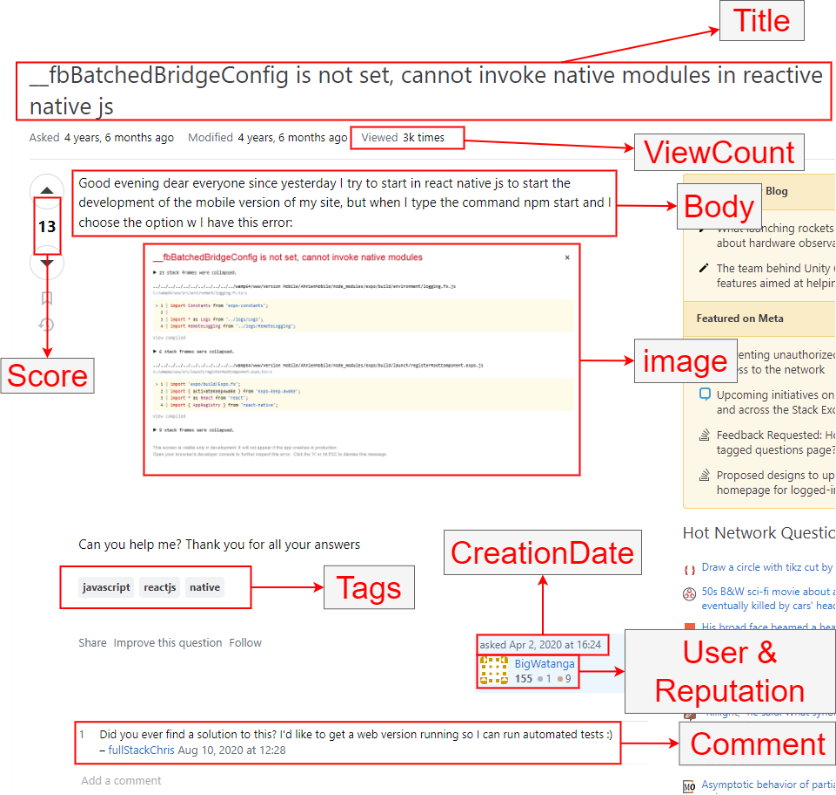}
    \caption{An example of a react-related discussion on Stack Overflow.}
    \label{fig:reactquestionexample}
\end{figure}

The data collection process was conducted through multiple stages.
The initial stage involved identifying several predictive attributes of React-related questions, as illustrated in~\cref{fig:reactquestionexample}.
In this step, we defined 12 attributes with their descriptions, as outlined in~\autoref{datasetattribute}.

\begin{table}[]
\caption{12 Extracted Attributes Used in This Study}
\resizebox{\columnwidth}{!}{%
    \begin{tabular}{|l|p{4cm}|p{4cm}|}
    \hline
    \textbf{Attribute} & \textbf{Description} & \textbf{Source} \\
    \hline
    Id & ID Number of the post & Extracted from data; used as unique identifier. \\
    CommentCount & Number of comments on a question post & Extracted from data; considered to reflect engagement. \\
    ViewCount & Number of times a specific question has been viewed by users & Extracted from data; included to measure visibility. \\
    Tags & Keywords or phrases associated with a question to categorize it & Extracted from data; excluded from final analysis. \\
    code\_snippet & Indicates if a block of code included within a question & Derived by detecting code within the post. \\
    question\_line\_count & Length of a question measured by their total number of lines & Calculated from question text. \\
    code\_line\_count & Number of lines of code included in the code snippet & Calculated from code snippets. \\
    image & Indicates if the question contains an image & Derived by detecting image tags. \\
    pd\_score & Performance difficulty score of the questions & Computed using a defined formula. \\
    Reputation & Reputation score of users who post the questions & Extracted from data linked by ID. \\
    ReputationCategory & Category of user reputation & Derived by grouping reputation scores. \\
    answered? & Indicates if the post has been answered & Determined by checking for answers. \\
    \hline
    \end{tabular}%
}
\label{datasetattribute}
\end{table}

The next step involved extracting data from the Stack Exchange Data Explorer archive,\footnote{\url{https://archive.org/download/stackexchange}} where it enables users to apply T-SQL (SQL variant) for various queries~\cite{hazoom2021text}.
By following prior research~\cite{kurniaji2023preliminary}, during the data collection process, we identified 23 relevant tags to extract React-related questions, as shown in~\autoref{freqpostseachtag}. The application of these tags resulted in 583,419 question posts, as presented in~\autoref{resultsdatacollection}.
We then conducted data preprocessing, which consisted of several stages. First, we removed duplicate questions, including those officially marked as duplicates by the moderation on SO, to reduce potential bias in answerability analysis. Next, we handled missing data by excluding records with missing essential fields (e.g., PostTypeId, Tags). Finally, we normalized and standardized attributes to ensure consistency across the dataset. These preprocessing steps resulted in 534,820 validated question posts related to React.

\begin{table}[]
\caption{Frequency of Posts for Each Tag Used in the Filtration}
\resizebox{\columnwidth}{!}{%
    \begin{tabular}{|c|l|r|c|l|r|}
    \hline
    \textbf{No}&\textbf{Tag}&\textbf{\#Posts}&\textbf{No}&\textbf{Tag}&\textbf{\#Posts} \\
    \hline
    1&reactjs&435,809&13&react-apollo&2,228 \\
    2&react-native&125,756&14&react-select&1,907 \\
    3&react-hooks&28,279&15&react-dom&1,068 \\
    4&react-redux&21,832&16&reactjs-flux&799 \\
    5&react-router&19,250&17&reactjs.net&170 \\
    6&react-navigation&8,032&18&konvajs-reactjso&106 \\
    7&create-react-app&5,494&19&reactjs-testutils&85 \\
    8&react-bootstrap&3,765&20&reactjs-native&49 \\
    9&react-testing-library&3,539&21&video-reactjs&22 \\
    10&react-props&3,037&22&reactjs-popup&8 \\
    11&react-native-ios&2,999&23&applicationinsights-react-js&5 \\
    12&react-native-flatlist&2,498& & &  \\
    \hline
    \end{tabular}%
}
\label{freqpostseachtag}
\end{table}

In extracting some attributes, such as \texttt{code\_snippet}, \texttt{code\_line\_count}, and \texttt{image}, we implemented regular expressions (RegEx) - a fundamental concept in formal language theory~\cite{chen2022solving}. The applied RegEx formats are as follows:

\begin{itemize}
    \item for code snippet detection
    
    \texttt{lambda x: 1 if re.search(r'\textless{}code\textgreater{}.* \textless{}/code\textgreater{}', str(x), re.DOTALL) else 0} 
   
    \item for code content extraction
    
    \texttt{re.findall(r'\textless{}code\textgreater{}(.*?)\textless{}/code\textgreater{}', str(text), re.DOTALL)} 
   
    \item for image element identification
    
    \texttt{lambda x: 1 if re.search(r'\textbackslash{}img\textbackslash{}s+ [\textasciicircum{}>\textbackslash{}]*src=', str(x)) else 0} 
\end{itemize}

\begin{table}[]
\caption{Number of Questions Resulted from Each Step}
\resizebox{\columnwidth}{!}{%
    \begin{tabular}{|l|p{5cm}|r|}
    \hline
    \textbf{Step} & \textbf{Description}    &   \textbf{\#SO Question Posts} \\
    \hline
    Step 1: & Initial raw data collection from SO & 24,243,707 \\
    Step 2: & Extraction of React-related questions using 23 specified tags & 583,419 \\
    Step 3: & Duplications removal & 534,820 \\
    \hline
    \end{tabular}%
}
\label{resultsdatacollection}
\end{table}

\subsection{Research Question}
To guide our research, we formulated two research questions along with their motivations and approaches:

\subsection*{\textbf{RQ$_1$: What attributes affect the probability of a React-related question on Stack Overflow being answered?}}

\textbf{Motivation:} The answerability of SO questions is an important variable to measure interaction quality on SO~\cite{chua2015answers}, especially React-related questions. Understanding the attributes associated with React-related question answerability helps developers create effective questions and provides benefits. For new developers, this understanding can improve their chances of getting answers. For experienced developers, it helps them to contribute more precisely to the community. Moreover, for SO platform managers, these findings can serve as a foundation for developing features that can guide users in writing higher-quality questions, leading to improve platform effectiveness.

\textbf{Approach:} To answer this RQ, we applied logistic regression with an $\alpha$ = 0.05 to analyze how individual question attributes relate to React question answerability.
This method was selected for its ability to predict binary responses with simple predictors and low incidence~\cite{nusinovici2020logistic}. 
For analyses where the dependent variable was binary (answered vs. unanswered), logistic regression is more appropriate for modeling categorical outcomes than ANOVA, which is typically used for continuous dependent variables. 
This technique is also frequently applied in software engineering research, such as estimating software development efforts~\cite{yang2022predictive} and the development of predictive models~\cite{hariyanti2024implementation}.
Taken from previous work~\cite{kleinbaum2002logistic}, the logistic regression model formula is presented in~\eqref{eq:logreg}.

\begin{equation}
    \label{eq:logreg}
    \text{Logit}\left(\frac{1}{1 + e^{-(\alpha + \sum\beta_i X_i)}}\right) = \ln\left(\frac{P(X)}{1 - P(X)}\right)
\end{equation}

\noindent\textbf{Description:}
\begin{itemize}
    \item $\alpha$: Constant or intercept
    \item $\beta_i$: Coefficient of the independent variable $X_i$
    \item $X_i$: Value of the independent variable
    \item $1 - P(X)$: Probability of an event not occurring
    \item $\ln$: Natural logarithm (base $e$)
\end{itemize}

We used the final dataset yielded from Step 3 as in~\autoref{resultsdatacollection}, by focusing on 7 selected important attributes, as presented in~\autoref{datasetattribute}, including \texttt{code\_snippet}, \texttt{image}, \texttt{Reputation}, \texttt{CommentCount}, \texttt{ViewCount}, \texttt{question\_line\_count}, and \texttt{code\_line\_count}. Next, we transformed these attributes into square root forms to address multicollinearity between \texttt{question\_line\_count} and \texttt{code\_line\_count}, which are interrelated, while the \texttt{Reputation} was transformed using the natural logarithm. Due to the imbalanced distribution between answered and unanswered questions, we applied SMOTE (Synthetic Minority Over-sampling Technique) only to the training dataset during model training to address class imbalance.

This analysis will provide insights into the relative contribution of each question attribute to the likelihood of receiving an answer.
Therefore, we propose the following hypotheses to test using the logistic regression model:

\begin{enumerate}[label=(\alph*)]
    \item H$_0$: Individual question attributes are not significantly associated with React question answerability.

    \item H$_1$: Individual question attributes are significantly associated with React question answerability.
\end{enumerate}

\subsection*{\textbf{RQ$_2$: Do user reputations correlate with the difficulty levels of React questions on Stack Overflow?}}

\textbf{Motivation:} User reputation is often assumed as an indicator of expertise within the SO community~\cite{wang2021reputation}. Understanding the correlation between user reputation and question complexity provides insights into user behavior and carries important implications for comprehending expertise dynamics. For new users, this insight can help meet expectations and increase confidence in participating. For developers in general, it can help create a more inclusive environment by understanding how users with different skill levels interact on the platform. For educators, it can help design more effective learning strategies.

\textbf{Approach:} In this RQ, a quantitative approach using linear regression with an $\alpha$ = 0.05 will be utilized.
For user reputation analysis, we categorized users into three classes based on their reputation scores. These categories follow the thresholds from Movshovitz-Attias et al. \cite{movshovitz2013analysis}, as presented in~\autoref{classificationuserreputation}.

\begin{table}[]
    \caption{Classification of User Reputation Based on Score~\cite{movshovitz2013analysis}}
    \begin{center}
    \begin{tabular}{|l|c|}
        \hline
        \textbf{Category} & \textbf{Score Interval} \\
        \hline
        High Reputation & $r \geq 2400$ \\
        Mid Reputation & $400 \leq r < 2400$ \\
        Low Reputation & $1 < r < 400$ \\
        \hline
        \multicolumn{2}{l}{$r = $ User score in SO.}
    \end{tabular}
    \label{classificationuserreputation}
    \end{center}
\end{table}

Next, we implemented a simple linear regression to analyze the correlation between reputation categories and question difficulty levels. 
Simple linear regression is suitable for isolating the correlation of independent attributes on quantitative dependent attributes and analyzing linear correlations~\cite{hope2020linear, james2023introduction, maulud2020review}. 
This technique is commonly applied in software engineering studies, such as evaluating the system usability tests~\cite{sukmasetya2020combining}. 
By following prior study~\cite{stolte2024comprehensive}, the simple linear regression model is presented in~\eqref{eq:linreg}.

\begin{equation}
    \label{eq:linreg}
    y = C + \beta x + \varepsilon
\end{equation}

\noindent\textbf{Description:}
\begin{itemize}
    \item $y$: Value to be predicted (dependent variable)
    \item $C$: Constant or intercept
    \item $\beta$: Regression coefficient
    \item $x$: Correlation factor $y$ (independent variable)
    \item $\varepsilon$: Error term (residual)
\end{itemize}

The question difficulty level was measured using a performance difficulty score (PD Score), which refers to a question complexity index, where lower scores indicate more complex questions.
The \textit{PD Score} considers the ratio between answer count and view count.
Based on prior research~\cite{islam2021network}, the \textit{PD Score} is calculated using~\eqref{eq:pdscore}.

\begin{equation}
    \label{eq:pdscore}
    \centering
    \textit{PD Score} = \frac{\textit{Average Answer Count}}{\textit{Average View Count}} \times 100%
\end{equation}

In this analysis, we proposed the following hypotheses to test using simple linear regression, focusing on the correlation between user reputation and \textit{PD Score}. 
Simple linear regression helps us understand if users' expertise level, reflected in their reputation, correlates with the complexity of questions they submit on SO.

\begin{enumerate}[label=(\alph*)]
    \item H$_0$: No correlation exists between user reputation and performance difficulty score.

    \item H$_1$: A correlation exists between user reputation and performance difficulty score.
\end{enumerate}

These hypotheses testing aims to prove whether high-reputation users tend to submit more complex questions, as demonstrated by low-performance difficulty scores. 

\begin{figure}
    \centering
    \begin{footnotesize}
    \begin{tikzpicture}
    \begin{axis}[
        xbar,
        align=left,
        width=.9\columnwidth,
        height=5cm,
        symbolic y coords={Answered, Unanswered},
        ytick=data,
        axis line style={opacity=0},
        major tick style={draw=none},
        y=0.7cm,
        xmin=0,
        xmax=100,
        bar width=15pt,
        nodes near coords={\pgfmathprintnumber[fixed,precision=2]\pgfplotspointmeta\%},
        nodes near coords align={horizontal}
    ]
    \addplot[fill=gray,draw=none] coordinates {
        (80.28,Answered)
        (19.72,Unanswered)
    };
    \end{axis}
    \end{tikzpicture}
    \end{footnotesize}
    \caption{Distribution of answered and unanswered questions.}
    \label{fig:answerdistribution}
\end{figure}
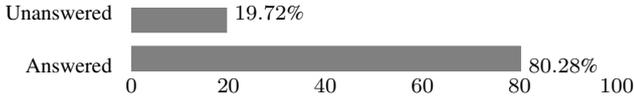

\subsection{Online Appendix}
\label{sec:online_appendix}
To facilitate the reproducibility of this study, we share our replication package online on~\url{https://github.com/vanesyaaura/SO-react-question-complexity}. 

\section{Results and Discussion}

\subsection{\textbf{RQ$_1$: What attributes affect the probability of a React-related question on Stack Overflow being answered?}}

To facilitate this RQ, we initially identify the distribution of answered and unanswered React-related questions shared on SO, as illustrated in~\cref{fig:answerdistribution}, which shows a significant difference between the number of answered and unanswered questions. 
Our preliminary result shows that out of 534,820 analyzed questions, 80.28\% received answers, while 19.72\% remained unanswered, demonstrating the high response of React-related questions from the community. To support subsequent classification tasks, SMOTE was applied to balance the dataset, mitigating the impact of class imbalance and enhancing the classifier's performance reliability.

\begin{figure}
    \centering
    \small
    \begin{tikzpicture}
        \begin{axis}[
            boxplot/draw direction=y,
            width=\columnwidth,
            height=4.75cm,
            ylabel={Number of Views},
            xtick={1,2},
            xticklabels={Unanswered, Answered},
            ymin=0, ymax=4500,
            ymajorgrids=true
        ]
        \addplot+[
            boxplot prepared={
                lower whisker=0,
                lower quartile=100,
                upper quartile=500,
                upper whisker=1100,
                median=185
            },
            fill=none
        ] coordinates {};
        \addplot+[
            boxplot prepared={
                lower whisker=0,
                lower quartile=200,
                upper quartile=1900,
                upper whisker=4400,
                median=640
            },
            fill=none
        ] coordinates {};
        \node[draw=none, fill=none, text=red, font=\small] at (axis cs:1,285) {Median: 185};
        \node[draw=none, fill=none, text=blue, font=\small] at (axis cs:2,740) {Median: 640};
        \end{axis}
    \end{tikzpicture}
    \caption{Number of views between answered and unanswered questions.}
    \label{fig:viewcountdistribution}
\end{figure}
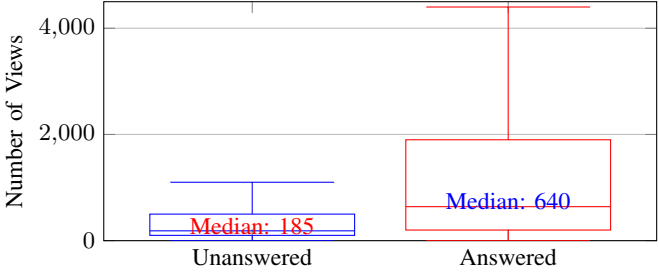

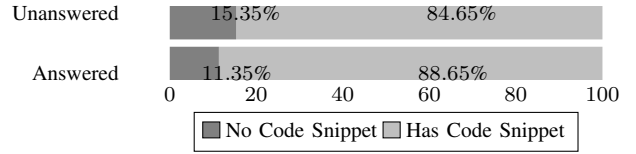
\begin{figure}
    \centering
    \footnotesize
    \begin{tikzpicture}
        \begin{axis}[
            xbar stacked,
            align=left,
            width=.95\columnwidth,
            height=7cm,
            symbolic y coords={Answered, Unanswered},
            ytick=data,
            axis line style={opacity=0},
            major tick style={draw=none},
            y=.8cm,
            xmin=0,
            xmax=100,
            bar width=20pt,
            nodes near coords={\pgfmathprintnumber[fixed,precision=2]\pgfplotspointmeta\%},
            nodes near coords align={horizontal},
            nodes near coords style={font=\footnotesize},
            legend style={at={(0.5,-0.45)}, anchor=north, font=\footnotesize},
            xlabel style={yshift=-20pt},
            enlarge x limits=0.1,
            legend columns=2,
            legend cell align={left}
        ]
        \addplot[fill=gray,draw=none] coordinates {
            (15.35,Unanswered)
            (11.35,Answered)
        };
        \addplot[fill=lightgray,draw=none] coordinates {
            (84.65,Unanswered)
            (88.65,Answered)
        };
        \legend{No Code Snippet, Has Code Snippet}
        \end{axis}
    \end{tikzpicture}
    \caption{Frequency of questions distinguished by their inclusion of code snippets.}
    \label{fig:codesnippetdistribution}
\end{figure}

In addition to our preliminary result, we also present the distribution of number of views, number of lines of code, and user reputation scores based on the availability of answers to React-related questions. As shown in~\cref{fig:viewcountdistribution}, the number of views is significantly different between answered and unanswered questions. It shows that more views seem to have higher chances of receiving answers from community, with its median of 640 views.
Although the inclusion of code snippets is not significantly different for both React-related questions with answers and those with no answers, as depicted in~\cref{fig:codesnippetdistribution}, it is still important to consider while posting the question.

\begin{figure}
    \centering
    \small
    \begin{tikzpicture}
        \begin{axis}[
            boxplot/draw direction=y,
            width=\columnwidth,
            height=4.75cm,
            ylabel={Number of Code Lines},
            xtick={1,2},
            xticklabels={Unanswered, Answered},
            ymin=0, ymax=16,
            ymajorgrids=true
        ]
        \addplot+[
            boxplot prepared={
                lower whisker=0,
                lower quartile=2,
                upper quartile=7,
                upper whisker=15,
                median=5.00
            },
            fill=none
        ] coordinates {};
        \addplot+[
            boxplot prepared={
                lower whisker=0,
                lower quartile=3,
                upper quartile=7,
                upper whisker=14,
                median=5.10
            },
            fill=none
        ] coordinates {};
        \node[draw=none, fill=none, text=red, font=\small] at (axis cs:1,5.5) {Median: 5.00};
        \node[draw=none, fill=none, text=blue, font=\small] at (axis cs:2,5.6) {Median: 5.10};
        \end{axis}
    \end{tikzpicture}
    \caption{Frequency of code lines included in both answered and unanswered questions. This chart is normalized using the square root to present the number of lines of code.}
    \label{fig:codelinecountdistribution}
\end{figure}
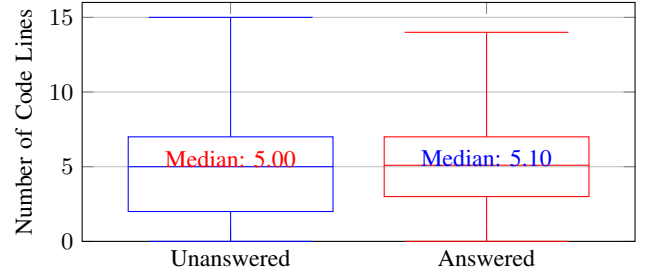

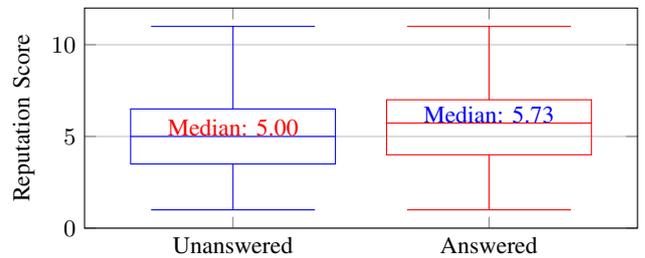
\begin{figure}
    \centering
    \small
    \begin{tikzpicture}
        \begin{axis}[
            boxplot/draw direction=y,
            width=\columnwidth,
            height=4.5cm,
            ylabel={Reputation Score},
            xtick={1,2},
            xticklabels={Unanswered, Answered},
            ymin=0, ymax=12,
            ymajorgrids=true
        ]
        \addplot+[
            boxplot prepared={
                lower whisker=1,
                lower quartile=3.5,
                upper quartile=6.5,
                upper whisker=11,
                median=5.00
            },
            fill=none
        ] coordinates {};
        \addplot+[
            boxplot prepared={
                lower whisker=1,
                lower quartile=4,
                upper quartile=7,
                upper whisker=11,
                median=5.73
            },
            fill=none
        ] coordinates {};
        \node[draw=none, fill=none, text=red, font=\small] at (axis cs:1,5.5) {Median: 5.00};
        \node[draw=none, fill=none, text=blue, font=\small] at (axis cs:2,6.2) {Median: 5.73};
        \end{axis}
    \end{tikzpicture}
    \caption{Distribution of user reputation scores for answered and unanswered questions. Due to its large variation in user reputation scores, we transformed it using a natural logarithm.}
    \label{fig:reputationdistribution}
\end{figure}

These findings are also applicable to the number of lines of code and user reputation scores.
As illustrated in~\cref{fig:codelinecountdistribution}, the number of code lines in the answered and unanswered questions is slightly different, where the questions with more lines of code suggesting to have higher probability of getting responses.
Furthermore, higher scores of user reputation also contribute to the chance of receiving replies from the community, as presented in~\cref{fig:reputationdistribution}.

\begin{table}[]
    \caption{Coefficients and P-Values of Each Testing Attribute Associated with the Answerability to React-related Questions}
    \begin{center}
    \begin{tabular}{|l|r|r|}
        \hline
        \textbf{Attributes} & \textbf{Coefficient} & \textbf{P-value} \\
        \hline
        code\_snippet & 0.38 & 0.00 \\
        image & -0.06 & 0.00 \\
        Reputation & 0.10 & 0.00 \\
        CommentCount & -0.40 & 0.00 \\
        ViewCount & 15.40 & 0.00 \\
        question\_line\_count & -0.35 & 0.00 \\
        code\_line\_count & 0.31 & 0.00 \\
        \hline
    \end{tabular}
    \label{resulttestingattributes}
    \end{center}
\end{table}

To complement our preliminary study, we describe the results of our hypothesis tests for 7 selected attributes, as shown in~\autoref{resulttestingattributes}. 
From the logistic regression implementation, it demonstrates that number of views, code snippets inclusion, number of code lines, and user reputation are positively associated with the answerability of a question, with positive coefficient of 15.40, 0.38, 0.31, and 0.10, respectively.
This indicates that the more frequently viewed questions, longer code snippets, and higher reputation of users may increase the probability of receiving answers.

In contrast, the presence of images, question length, and number of comments provide negative associated with coefficients of -0.06, -0.35, and -0.40 respectively. 
This means that fewer images, question lengths, and comments may increase the chance of getting an answer, and suggests that excessive elements can reduce the probability of receiving answers, potentially due to reduced efficiency and focus, which may hinder productive community interaction.

Comparing our findings with a prior study on Python-related questions~\cite{nugroho2024empirical} reveals similar patterns. Both studies confirm user reputation as a key factor in question answerability across domains. Our results also reflect themes from prior taxonomies of unanswered questions~\cite{asaduzzaman2013answering}. For example, React questions with excessive text, unclear phrasing, or missing code align with categories like ``too vague'' or ``no code.'' Technically complex or narrowly focused questions may be considered ``too time-consuming'' or ``fail to attract experts,'' where answers are lacking despite clarity. These parallels highlight the general importance of clarity, context, and code inclusion in improving answerability on Stack Overflow.

\begin{table}[]
    \caption{Frequency of Question Posts for Each Category of User Reputation}
    \begin{center}
    \begin{tabular}{|l|r|r|}
        \hline
        \textbf{Reputation Category}&\textbf{\#Questions}&\textbf{Percentage} \\
        \hline
        Low &   302,977 &   56.65\% \\
        Mid &   146,409 &   27.38\% \\
        High    &   85,434  &   15.97\% \\
        \hline
    \end{tabular}
    \label{freqpostseachreputationcategory}
    \end{center}
\end{table}

\subsection{\textbf{RQ$_2$: Do user reputations correlate with the difficulty levels of React questions on Stack Overflow?}}

Before we further analyze the relationship between user reputation categories and performance difficulty scores, we initially classified the users' category who posted React-related questions on SO based on their reputation scores.
As presented in~\autoref{freqpostseachreputationcategory}, the React-related questions were mostly posted by the low-reputation users, accounting for 56.65\%, followed by the mid and high-level users, as many as 27.38\% and 15.97\%, respectively. This illustrates to the community that React-related issues are usually a challenge for SO newcomers.

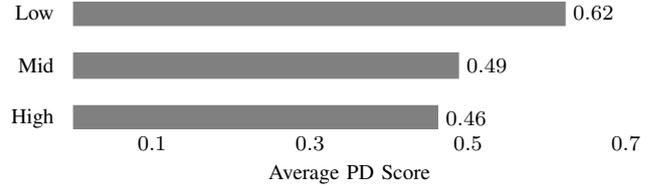
\begin{figure}
    \centering
    \begin{footnotesize}
    \begin{tikzpicture}
    \begin{axis}[
        xbar,
        align=left,
        width=\columnwidth,
        height=4cm,
        symbolic y coords={High, Mid, Low},
        ytick=data,
        axis line style={opacity=0},
        major tick style={draw=none},
        y=0.7cm,
        xmin=0,
        xmax=.7,
        xlabel={Average PD Score},
        nodes near coords,
        nodes near coords align={horizontal},
        bar width=10pt,
        xtick={0.1, 0.3, 0.5, 0.7}
    ]
    \addplot[fill=gray,draw=none] coordinates {
        (0.623673,Low)
        (0.488686,Mid)
        (0.462403,High)
    };
    \end{axis}
    \end{tikzpicture}
    \end{footnotesize}
    \caption{Average difficulty score of questions shared by each category of users.}
    \label{fig:averagepdscore}
\end{figure}

\begin{figure}
    \centering
    \includegraphics[height=.6\columnwidth]{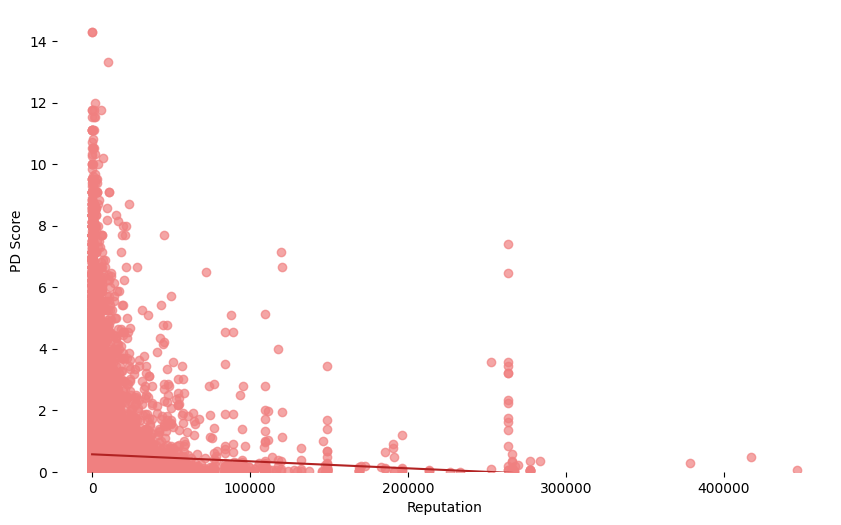}
    \caption{Correlation between reputation and \textit{PD Score}.}
    \label{fig:correlation}
\end{figure}

Furthermore, in this study, we also calculated the average PD scores of each category of user reputation.
As shown in~\cref{fig:averagepdscore}, low-reputation users occupy the top spot in the average PD score, at 0.62, which highlights the less difficult questions posted by them.
In contrast, high-reputation users have the most difficult React-related problems in SO compared to other user reputation categories, as indicated by the lowest average PD score of 0.46.
This analysis is further supported by~\cref{fig:correlation}, which shows a clear negative trend between PD scores and user reputation.
This implies that users with higher reputations tend to submit more challenging questions on SO.

To strengthen our experiments, we further investigated the assumption in the proposed hypothesis by implementing a simple linear regression model.
The hypothesis test shows a negative correlation between user reputation scores and performance difficulty scores of questions, with a regression coefficient of -0.092 and an intercept of 0.561, as described in~\autoref{tab:linreg_parameter}. 
This significant negative correlation suggests that users with high credentials on SO, as reflected by their user reputations, possess the cognitive capacity and technical abilities that enable them to formulate questions with more advanced difficulty and specificity.

\begin{table}[]
    \centering
    \caption{Coefficient and Intercept from the Implementation of Simple Linear Regression to User Reputations and PD Score}
    \begin{tabular}{|l|r|}
        \hline
        \textbf{Parameter} & \textbf{Value} \\ 
        \hline
        Slope/Coefficient  & -0.092    \\  
        Intercept          & 0.561  \\  
        \hline
    \end{tabular}
    \label{tab:linreg_parameter}
\end{table}

In sum, these results provide insights for developers in creating effective React-related questions on Stack Overflow, highlighting the attributes that increase the likelihood of receiving answers from the community. Beyond individual benefits, understanding the factors associated with question answerability can improve knowledge exchange on Q\&A platforms. Our findings can inform educational resources, guide community moderation, support recommendation systems to help users craft more answerable questions, and ultimately improve knowledge sharing within the software development community.

\section{Threats to Validity}
Several potential threats to validity exist in our research. The \textbf{internal validity} concerns using the most recent user reputation scores rather than scores at question posting time, potentially affecting our analysis. 
Moreover, regarding RQ2, we focused solely on examining the relationship between user reputation and PD Score as a case study. While broader correlation analysis across all features could offer further insights, we leave this as a direction for future work.

\textbf{The threat to construct validity} exists in our view count analysis. Our cross-sectional data cannot determine causality direction—higher view counts may result from answered questions rather than causing answers, as questions with quality answers may attract more traffic.

The \textbf{external validity} relates to generalizability, as we focus specifically on the SO platform and React framework. We mitigate \textbf{threats to reliability} through our online dataset appendix (see Section~\ref{sec:online_appendix}).

\section{Conclusion}
This study investigates attributes associated with React-related questions' answerability and complexity levels. Results show that: (i) view counts (though causality requires further investigation), code snippet inclusions, code line numbers, and user reputations may increase answerability, while excessive images, question length, and comments potentially reduce it; and (ii) higher reputation users tend to post more complex React-related questions.

These findings establish a baseline for future work analyzing JavaScript framework question acceptability, exploring temporal trends, developing longitudinal analyses to clarify the causal relationship between views and answerability, and creating real-time predictive tools to assess these factors.


\balance

\end{document}